\documentclass[article]{has40}
\usepackage{graphicx}
\usepackage{amssymb}
\tabletypesize{\normalsize}  % AMcW

\shortauthors{Spalding, Wilhelm, \& De Lee}
\shorttitle{New Metallicity Calibration}

\begin{document}
\large    %AMcW  The conference proceedings will employ large size print
\pagenumbering{arabic}
\setcounter{page}{198}

\title{A New RR Lyrae Metallicity Calibration
Including High-Temperature Phase Regions}

%
% Here is an example of how to include the author names and affiliations
%
\author{{\noindent E. Spalding,{$^{\rm 1}$} R. Wilhelm,{$^{\rm 1}$} N. De Lee{$^{\rm 2}$}\\
\\
{\it (1) University of Kentucky, Lexington, KY, USA\\
(2) Vanderbilt University, Nashville, TN, USA} 
}
}

%
% And here is how to add the e-mail addresses
%
\email{(1) espalding@uky.edu, ron.wilhelm@uky.edu (2) nathan.delee@vanderbilt.edu}

%% Are the above institutional/email footnotes correctly formatted?
%% (ES and RW are from institution (1) and NDL from institution (2); ES
%% email is (1), RW email is (2), NDL email is (3))

\begin{abstract}
We have begun developing a new metallicity calibration for RR Lyrae
stars that includes high-temperature phase regions. Our calibration is
an updated and expanded version of the Delta-S, equivalent width
method, but which is applicable for most phases of the pulsation, and
for the hotter, shorter-period RRc stars. This calibration will be
constructed from spectral observations of well studied, bright RR
Lyrae stars with metallicities from published high-resolution studies,
within the range $-2.5 < [Fe/H] < -0.3$. Phase information for our
calibration stars were taken from the AAVSO website, and the MacAdam
Student Observatory at the University of Kentucky. Low-resolution
spectroscopy $(R \sim 1000)$ was obtained using the McDonald
Observatory, 2.1 meter telescope. Here we present our preliminary
data, and find the phase region where the calibration will break
down. 
\end{abstract}

\section{Introduction}
% first paragraph
Previous efforts at RR Lyrae metallicity calibrations have tended to
shy away from using spectroscopic observations near maximum light (an
exception is Butler 1975), mostly in order to avoid the NLTE effects
of shock waves in the atmosphere (e.g., Clementini et al. 1995,
Layden 1994, Fernley \& Barnes 1996, Lambert 1996). Furthermore,
RRc’s, which tend to be hotter than RRab’s, are underrepresented in
the literature. Here we aim to extend a metallicity calibration to the
high-temperature regions of both RRab’s and RRc’s, and determine where
precisely the calibration fails.

% second paragraph
RR Lyrae have already been used to help trace out streams and
regions of the galactic halo (e.g., Layden 1995, Drake et
al. 2012). Since 1998, a dedicated telescope in New Mexico has been
mapping the night sky for the Sloan Digital Sky Survey\footnote{http://www.sdss3.org/dr9} (SDSS), 
which currently includes a data set of more than 660,000 stellar
spectra. Ultimately, we hope to use
our new calibration to determine RR Lyrae metallicities from
multi-epoch spectra in the SDSS database. 

\section{Observations}
% first paragraph
RR Lyrae stars with published periods and metallicities determined
from high-resolution spectroscopy were selected for observation. Stars
were observed at the McDonald Observatory, 2.1 meter telescope with
the ES2 spectrometer $(R \sim 1000)$ in December 2012, with another
run planned for July 2013. Spectra were acquired to attain as full a
phase coverage as possible, with exposure times ranging from 300 to
1800 seconds. Exposures were limited to less than 10\% of the phase to
avoid averaging over phase changes. Spectra were reduced and
normalized in IRAF, and equivalent widths were measured manually. 

% second paragraph
In order to determine the time of last maximum, light curves were
initially sought on the AAVSO
website\footnote{http://www.aavso.org/data/lcg}. If a light curve was
available, the last recorded maximum was used to predict the time to
observe the star at the MacAdam Student Observatory on the University
of Kentucky campus. Light curves were taken at MacAdam, primarily in
December 2012 and January 2013, with a time resolution of less than 30
seconds. Fourth-order polynomials were fit to the light curves, which
yielded a maximum with an error of less than ten minutes.  

\section{Preliminary Results}
\begin{figure}
\centering
\includegraphics[width=\textwidth]{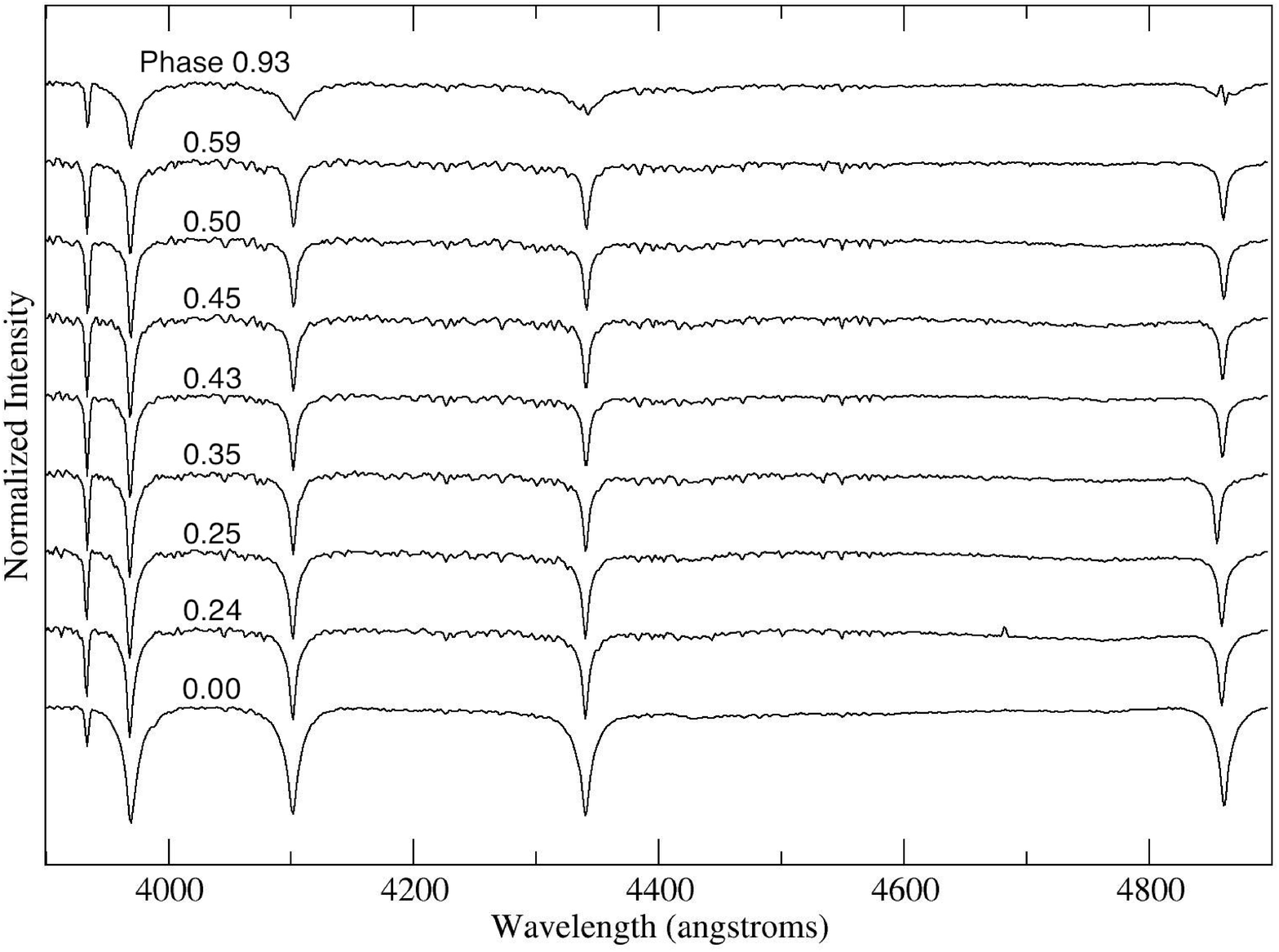}
%\vskip 0pt
\caption{Offset spectra of X Ari at different phases.}
\label{phase}
\end{figure}

\begin{figure}
\centering
\includegraphics[width=\textwidth]{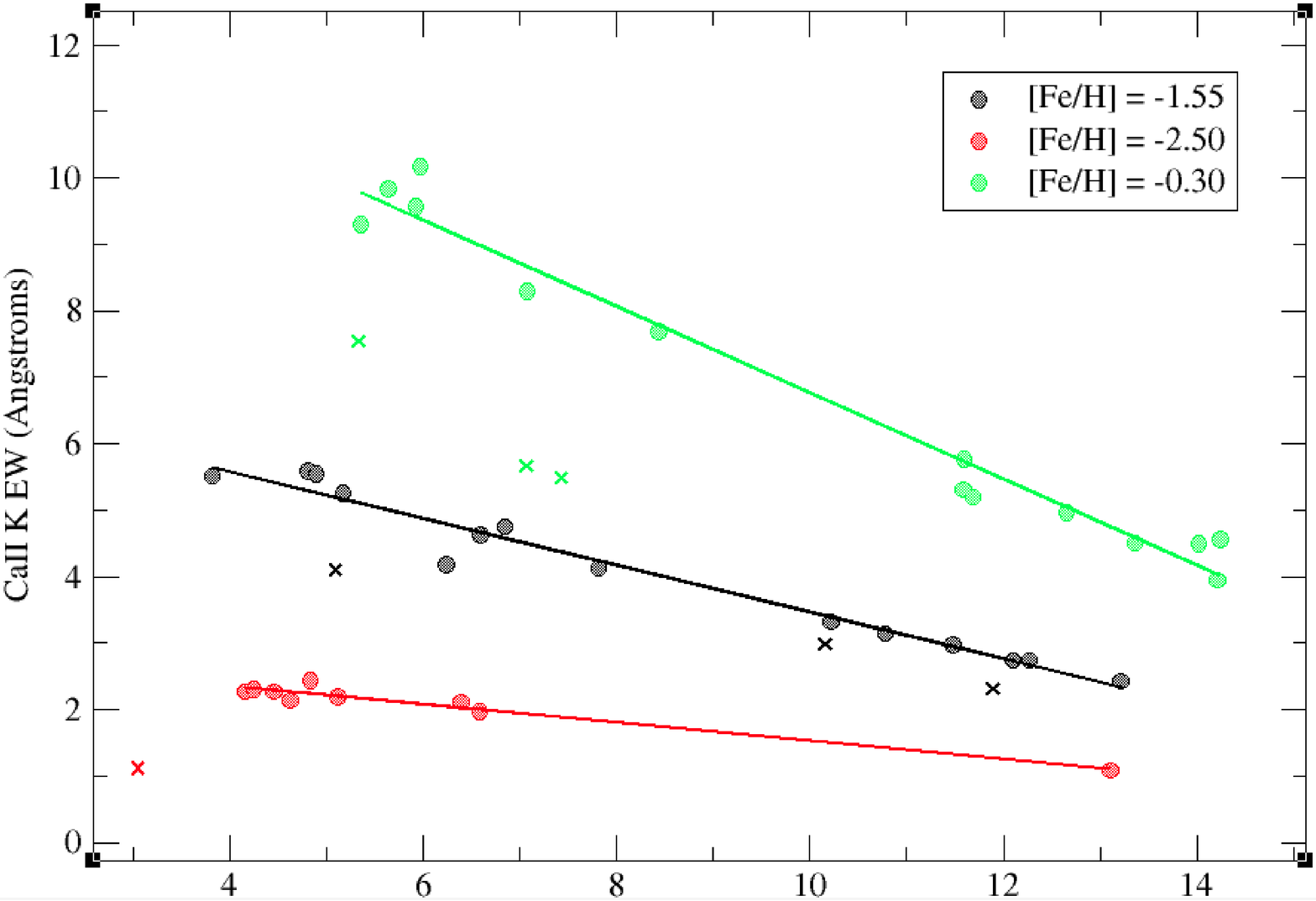}
%\vskip 0pt
\caption{A preliminary view of the CaII-K and hydrogen
  line width trend. Excluded points $(0.85 < p < 0.93)$ are crosses. The lower line
  corresponds to X Ari, and the upper line to AR Per. The line in between is a combination of TT Lyn and TU UMa.}
\label{trends}
\end{figure}

We recorded spectra at multiple phases for 14 RR Lyrae stars. The
effect of possible shock waves (Chadid et al. 2008) is evident
at phase 0.93 in Figure \ref{phase}, where we show the normalized
spectrum of X Ari. The CaII-K and hydrogen equivalent width trends of
four stars are plotted in Figure \ref{trends}. The hydrogen lines are
an average of the $\beta$ through $\delta$ lines, where the $\beta$
and $\gamma$ lines have been stretched to match the corresponding
$\delta$ line. It can be seen that the widths maintain a fairly linear
relationship if the phase regions $0.85<p<0.93$ are excluded. The
excluded data points (crosses) are likely where NLTE effects
associated with shock waves are present. 

\pagebreak 

\section{Acknowledgments} Thanks go to Kyle McCarthy for assistance
during observing runs at MacAdam, and Tim Knauer for his spreadsheet program.

\end{document}